\documentclass[twocolumn,showpacs]{revtex4}

\usepackage{graphicx}%
\usepackage{dcolumn}%
\usepackage{amsmath}%

\begin{document}



\title{Enhanced paramagnetism of the 4$d$ itinerant electrons in the rhodium
oxide perovskite SrRhO$_3$}

\author{K. Yamaura}
\email[E-mail at:]{YAMAURA.Kazunari@nims.go.jp}
\affiliation{Advanced Materials Laboratory, National Institute for Materials
Science, 1-1 Namiki, Tsukuba, Ibaraki 305-0044, Japan}
\affiliation{Japan Science and Technology Corporation, Kawaguchi, Saitama 
332-0012, Japan}

\author{E. Takayama-Muromachi}
\affiliation{Advanced Materials Laboratory, National Institute for Materials
Science, 1-1 Namiki, Tsukuba, Ibaraki 305-0044, Japan}

\begin{abstract}
Polycrystalline rhodium(IV) oxide perovskite SrRhO$_3$ was obtained by 
high-pressure synthesis techniques, followed by measurements of the magnetic
susceptibility, electrical resistivity, and specific heat.
The title compound has five 4$d$-electrons per perovskite unit and shows 
Fermi-liquid behavior in its electrical resistivity.
The magnetic susceptibility is large [$\chi({\rm 300K}) \sim1.1\times10^{-3}
$ emu/mol-Rh] and proportional to $1/T^2$ ($<$ 380 K), while there is no
magnetic long-range order above 1.8 K.
The specific heat measurements indicate a probable magnetic contribution below 
$\sim$ 15 K, which is not predicted by the self-consistent 
renormalization theory of spin fluctuations for both antiferro- and
ferromagnetic 3D nearly-ordered metals.
\end{abstract}



\maketitle

\section{Introduction}
Since $p$-wave symmetrical pairing of electrons was proposed, driven mainly by 
ferromagnetic spin fluctuations, in the 1.5 K superconductor Sr$_2$RuO$_4$ 
\cite{PT01YM}, further superconducting phases have been expected in the 
vicinity of the 214 phase. 
This is probably due to substantial spin fluctuations found in neighboring compounds,
including ferromagnetic SrRuO$_3$ \cite{PRB99JO,PRL99KY}, and nearly 
ferromagnetic CaRuO$_3$ \cite {PRL99KY,PRB01TH} and Sr$_3$Ru$_2$O$_7$ \cite
{PRB00SII}.
Although intensive investigations have been applied to the ruthenium oxide 
systems, further `$p$-wave' superconducting phases have not been discovered 
thus far.
The current experimental studies on ferromagnetically induced 
superconductivity, then, seem to be tied to a very local variety of materials. 
To ameliorate the stagnant situation, we have been exploring other correlated
$4d$-metal compounds, not only to find further superconducting materials in the 
ruthenium oxide system, but also to expand the variety of potential chemical 
systems for the spin-fluctuations-induced superconductors. 

The rhodium oxide perovskite SrRhO$_3$ was recently found, and a pure
polycrystalline sample was obtained by high-pressure synthesis techniques
at 60 kbar and 1500 $^\circ$C, followed by investigations of the magnetic 
susceptibility, electrical resistivity, and specific heat.
The compound was fairly metallic and showed enhanced and thermally activated 
paramagnetism in the studied temperature range below 380 K.
A qualitative fit of the Curie-Weiss (CW) law to the magnetic susceptibility
data yielded a negative Weiss temperature of -361 K, if the analysis provided
a correct sense of the magnetism.
Neither superconductivity nor long range magnetic order was found above 1.8 
K.
The magnetic data for SrRhO$_3$ appeared to be qualitatively similar to what
was observed for the analogous ruthenium oxide metal CaRuO$_3$ \cite 
{PRL99KY,PRB01TH}.
Since then, the self-consistent renormalization (SCR) theory of spin 
fluctuations for both antiferro- and ferromagnetic nearly-ordered magnetic 
metals was tested on the observed electronic properties as was done for 
CaRuO$_3$ \cite{PRL99KY,PRB01TH}.
As a result, all of the present data for SrRhO$_3$ do not meet the 
quantitative expectations as predicted by the theory.

\section{Experimental}
Variable composition precursors were prepared at Sr:Rh = 1:3, 1:2, and 1:1 as
follows.
Mixtures of pure SrCO$_3$ (99.9 \%) and Rh (99.9 \%) powders were heated in 
oxygen at 1000 $^\circ$C overnight, and then ground well and reheated in 
oxygen at 1200 $^\circ$C for two days \cite{binarysystem}.
One and two moles of SrO$_2$ ($>$99.9 \%) were added to the 1:2 and 1:3 
precursors per the formula, respectively, and 8 wt.\% of KClO$_4$ to the 1:1
precursor.
Those were mixed well, and approximately 0.2 g of each were placed into Pt 
capsules.
Those were heated at 60 kbar and 1500 $^\circ$C for 1 hr, then quenched to 
room temperature before releasing the pressure \cite{press}.
Quality of the finally obtained pellets was studied by powder x-ray-diffraction 
techniques in a regular manner.
It was determined from the x-ray readings that the major phase was of 
perovskite-type.
The position and intensity distribution of the peaks for the phase were 
invariable among the patterns for every sample.
The impurity level was 1 \% or less in every final production except KCl.
The perovskite-type phase denoted SrRhO$_3$, of which no records were 
found thus far in the literature. 

\begin{figure*}
\includegraphics{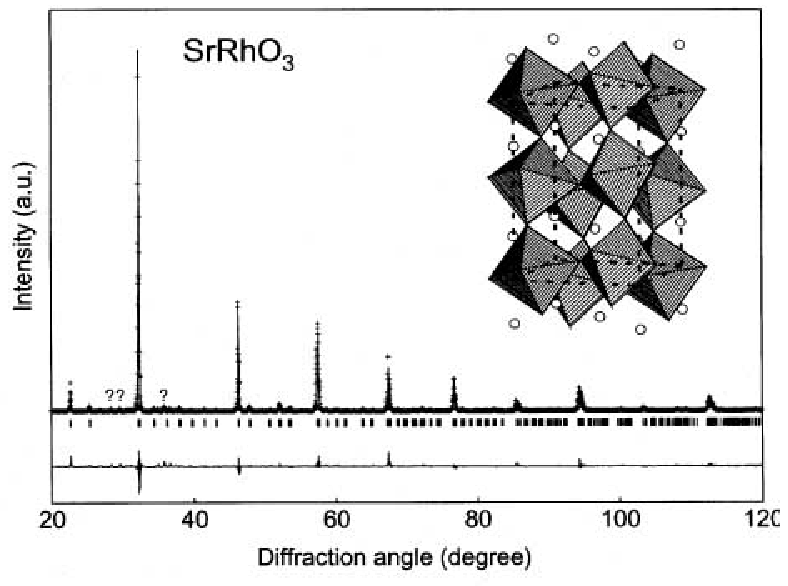}
\caption{The powder X-ray diffraction pattern (Cu$K\alpha$) for SrRhO$_3$. The
vertical bars show the Bragg peak positions for SrRhO$_3$. The difference 
plot between the orthorhombic model pattern (solid lines) and the data 
(crosses) is shown below the bars. Unknown peaks are marked by `?'. The 
crystal structure sketch with the orthorhombic unit cell (dotted lines) is 
shown as an inset. Open circles and polyhedra indicate Sr atoms and RhO$_6$ 
octahedra, respectively.}
\label{XRD}
\end{figure*}

Further structural characterization was made for the selected sample, which was
prepared from the 1:2 precursor and SrO$_2$, by x-ray Rietveld technique 
(Cu$K\alpha$) using the program RIETAN-2000 \cite{Rietveld}.
A distorted perovskite structure model, GdFeO$_3$-type, was tested and found
reasonable to describe the structure of SrRhO$_3$.
The x-ray powder pattern and crystal structure are 
indicated in Fig.\ref {XRD}; Space group was 
$Pnma$ (no. 62) and lattice parameters were $a =$ 5.5394(2) \AA, $b =$ 7.8539
(3) \AA, and $c =$ 5.5666(2) \AA.
The estimated positions for the atoms were Sr(0.0304(1), 0.25, -0.0054(8)), 
Rh(0, 0, 0.5), O1(0.4990(23), 0.25, 0.0587(45)), and O2(0.2825(26), 0.0366(24)
, 0.7088(26)).
During the refinement, the occupation factors, and the isotropic displacement
parameters of the metals and oxygen were fixed at 1, 0.3, and 0.7, 
respectively. 
The final reliability factors and goodness of fit to the analysis were
$R_{\rm wp} =$20.9 \%, $R_{\rm p} =$14.41 \%, $R_{\rm R} =$18.57 \%, and $S 
=$ 1.53.
Oxygen vacancies in the perovskite were quantitatively investigated in detail
by thermogravimetric analysis and found to be insignificant \cite{TGA}.

The same sample was again selected for characterization by magnetic, specific
heat, and electrical resistivity measurements.
The temperature dependence of magnetization was measured in a Quantum Design
MPMS magnetometer.
The specific heat and the electrical resistivity data were obtained in a 
Quantum Design PPMS apparatus.
Those measurements were conducted between 1.8 and 400 K. 
The highest applied magnetic field was 70 kOe.

\section{Results and Discussions}
\begin{figure*}
\includegraphics{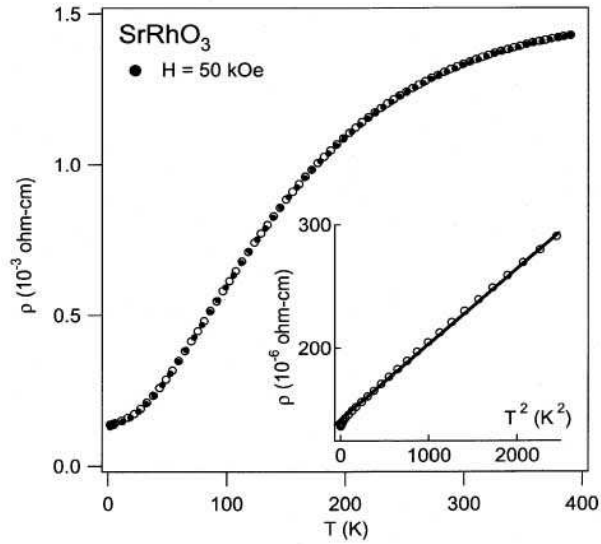}
\caption{The resistivity data for the polycrystalline SrRhO$_3$ measured 
between 1.8 and 390 K with and without a magnetic field of 50 kOe. Squared 
temperature plot for the data below 50 K is shown in the inset, indicating a
linear dependence as shown by the solid line.}
\label{resistivity}
\end{figure*}

Temperature dependence of the electrical resistivity of polycrystalline SrRhO$_3$
is shown in Fig.\ref{resistivity}. 
The data were obtained by a standard 4-terminal dc technique at a
current of 5 mA on a piece of the sample pellet.
The data clearly reveal the metallic nature of SrRhO$_3$; a metallic temperature 
dependence and $\sim$ 1.3 m$\Omega$cm at room temperature are typical for
polycrystalline oxide metals.
The low temperature part ($<$50 K) is expanded and replotted as $\rho$
vs $T^2$ (inset in Fig.\ref {resistivity}).
The observed linear dependence is indicative of Fermi liquid behavior for 
SrRhO$_3$ \cite{RMP98MI}. 
Subsequent fitting studies with standard resistivity expression for a Fermi liquid
($\rho = \rho_0 + AT^2$) yielded $\rho_0 =$ 142 $\mu\Omega$cm and $A =$ 0.062 
$\mu\Omega$cm/K$^2$.
The unusually large $\rho_0$ probably reflects contributions from 
extrinsic origins such as grain boundaries.
The parameter $\rho_0$ was not constant among the sets of resistivity data for
all of the present pellets (approximately two magnitudes larger for the pellet
containing KCl), while the residual resistivity ratio, $\rho_{300}/\rho_0$, 
remained almost constant ($\sim$ 9) among them. 
At the magnetic instability point, or in the extreme vicinity of that point, the electrical
resistivity is not expected to obey the famous $T^2$ law due to the influence of spin 
fluctuations; i.e. $T^{3/2}$ and $T^{5/3}$ law may be obeyed by antiferro- and
ferromagnetically unstable 3D metals, respectively \cite 
{JPCM96SRJ,PRB93AJM,JPSJ98AI}.
Detailed analysis was preliminarily applied for the present resistivity data,
however, the non-Fermi liquid behavior was not clearly seen.
Magnetoresistivity at 1.8 K between -70 and 70 kOe was not observed, and may
be due to polycrystalline nature of the sample.
Additional studies using a single crystal SrRhO$_3$, if it becomes available, could 
allow us to exclude the extrinsic contributions and then might help to reveal the 
intrinsic nature of electrical resistivity of SrRhO$_3$.
Because the 4$d$-band in SrRhO$_3$ is expected to be broad, as is the case for 
SrRuO$_3$, 4$d$-electrons in the rhodium oxide should be itinerate by analogy 
\cite{PRB99JO}.
The observed metallic conductivity is, hence, reflecting mainly the 
nature of unlocalized 4$d$ electrons.
The perovskite SrRhO$_3$ could be in a class of the itinerant 4$d$-electron
systems, such as (Sr,Ca)RuO$_3$ \cite{PRB99JO}.
Further investigations into the electronic transport of SrRhO$_3$, including band
structure calculations, would be of interest. 

\begin{figure*}
\includegraphics{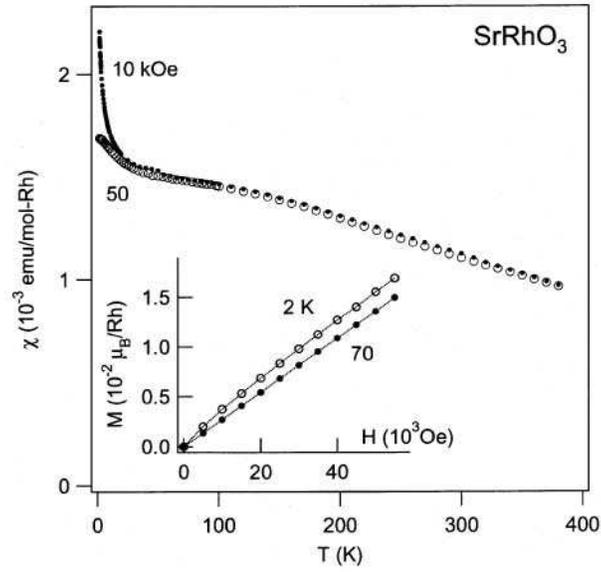}
\caption{The magnetic data for the polycrystalline SrRhO$_3$ measured between
1.8 and 400 K. Magnetic susceptibility vs temperature at 10 and 50 kOe are 
shown as closed and open circles, respectively, and the field dependence of 
magnetization at 2 and 70 K (inset).}
\label{magnetic}
\end{figure*}

Magnetic data are summarized in Fig.\ref{magnetic}. 
The magnetic susceptibility of SrRhO$_3$ obviously depends on temperature and is 
approximately 1.1$\times10^{-3}$ emu/mol-Rh at room temperature, in contrast 
with the properties of the Pauli paramagnetic rhodium metal (approximately one 
magnitude smaller and almost temperature independent) \cite{PPS61HK}.
A steep rise in the $\chi$ vs $T$ plot in low temperature at 10 kOe was observed,
while it was significantly suppressed at 50 kOe. 
The $M$ vs $H$ curve at 2 K (inset in Fig.\ref{magnetic}) indicates a subtle
spontaneous magnetic moment ($\sim$ 0.001 $\mu_{\rm B}$ per Rh), suggesting SrRhO$_3$
has ordered magnetic moments.
After subtraction of the major part, $1/\chi_{\rm upturn}$ vs $T$ plot results
in a standard CW line with an insignificant level of Weiss temperature 
$\sim$ -1.5 K \cite{upturn}.
It is therefore reasonable to conclude that the upturn results from a magnetic
impurity origin rather than an ordered state of SrRhO$_3$.
The slightly positive curvature of the $M$ vs $H$ curve at 2 K is probably due
to superimposing the small amount of impurity component on the major part.

\begin{figure*}
\includegraphics{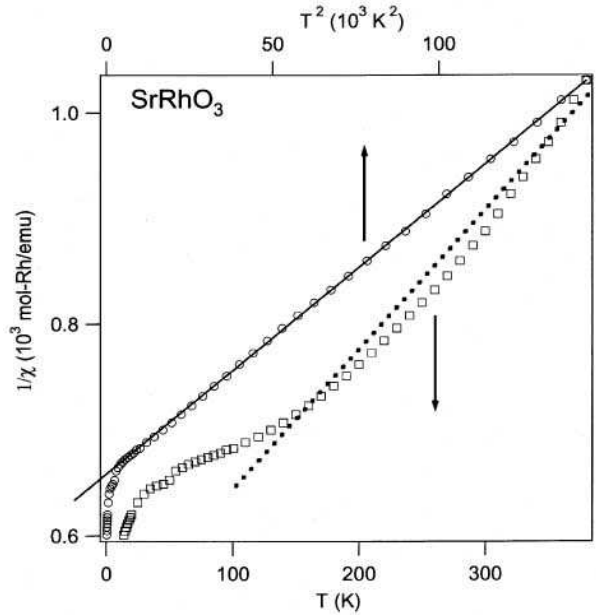}
\caption{Variable plotting of the reciprocal magnetic susceptibility of 
SrRhO$_3$. Contribution from the sample holder was carefully subtracted before the
plotting. Dotted line represents preliminarily applied CW law to the $1/\chi$
vs $T$ plot at $p_{\rm eff} =$ 2.46 $\mu_{\rm B}$ and $\theta_{\rm W} =$ -361
K. The $1/\chi$ vs $T^2$ plot shows a notably linear dependence as the solid line
indicates.}
\label{magnetic2}
\end{figure*}

To further analyze the major part of the magnetic data for SrRhO$_3$, two plots 
of the reciprocal magnetic susceptibility were prepared in the forms of $1/\chi$
vs $T$ and $1/\chi$ vs $T^2$ without any other manipulations except
subtraction of sample holder contribution (Fig.\ref{magnetic2}). 
It is clearly seen in the temperature range that $1/\chi$ is proportional to
$T^2$ rather than proportional to $T$ as expected from the standard CW 
expression.
Alternatively, the CW law with a temperature-independent term, i.e. $1/\chi 
= 1/[C/(T-\theta)+\chi_0]$ ($C$ and $\theta$ are the Curie constant and Weiss 
temperature, respectively), was applied to the $1/\chi$ vs $T$ plot.
The fit, however, failed to produce a convincible result \cite{preCW}. 
Tentative CW parameters obtained in the calculations were considerably sensitive to
least squares fitting conditions, including temperature range width, and stable 
and reasonable solutions were never found.
Further attempts were made to demonstrate the implied linear relationship between
$1/\chi$ and $T^2$ for SrRhO$_3$.
Neither the $T^{3/2} $ nor the $T^{4/3}$ fit (data not shown), however,
yielded a linear part, which was expected, if SrRhO$_3$ was just at
the magnetic instability point \cite{PRB93AJM,JPSJ98AI}. 
The above experimental observations would suggest that the magnetic 
susceptibility for SrRhO$_3$ is rather uncommon among properties of 
magnetic metals, because 
many antiferro- and ferromagnetic metals are expected to follow 
approximately the CW law above the magnetic ordering temperature or 0 K (in the case
for nearly ordered metals) \cite
{JPCM96SRJ,PRB93AJM,JPSJ98AI,Springer85TM,JPSJ73TM,JPSJ75HH,JPSJ87RK}.
The roughly estimated $\chi(0) \sim 1\times10^{-3}$ cm$^3$/mol for SrRhO$_3$,
and the Sommerfeld constant discussed later ($\gamma = 7.6$ mJ/mol K$^2$), 
yielded the Wilson ratio ($R_{\rm W}$) of $\sim$ 8.6 using the formula \cite{RMP75KGW},
\begin{eqnarray}
R_{\rm W} = {3\pi^2k_{\rm B}^2\chi(0) \over \mu_{\rm B}^2\gamma}.
\label{RW} 
\end{eqnarray} 
The preliminary $R_{\rm W}$ for SrRhO$_3$ is clearly out of the expected range, 1 
to 2, for standard Fermi-liquid behavior.
The unreliable $R_{\rm W}$ might support the presence of peculiar 
magnetism in SrRhO$_3$.

The most advanced profiling thus far achieved for the nearly and weakly 
antiferro- and ferromagnetic 3D metals was accomplished by developing the SCR
theory of spin fluctuations in metals \cite{Springer85TM}.
At the paramagnetic region, $1/\chi$ is expected to be in direct proportion 
to the $d$th power of $T$, where $d=$1 to 3/2 and 1 to 4/3 for antiferro- and
ferromagnetic 3D metals, respectively \cite{JPSJ98AI}.
This is the most notable point to distinguish the progress of 
understanding in magnetism of metals achieved by the SCR studies, and so-far
observations, indeed, seem to be in the range ($1/\chi \sim T^d$) \cite{JPSJ73TM}. 
The rather conventional Stoner's model ($1/\chi \sim T^2$) is far beyond the range.
The rhodium oxide metal, however, shows a nearly $T^2$
dependence of $1/\chi$, which ironically matches the Stoner expectation
\cite{Springer85TM}.
Although the $T^2$ trend in $1/\chi$ was also predicted by a random phase 
approximation theory, using it here to analyze the present data may be 
problematic, because it is too limited in temperature range (only effective 
within extremely low temperature), due to mainly a lack of self consistency 
\cite{Springer85TM,PRL68MTB}. 
Further considerations with additional studies may be necessary to 
conclusively determine the microscopic origin of the $1/\chi \sim T^2$ trend
in SrRhO$_3$.

\begin{figure*}
\includegraphics{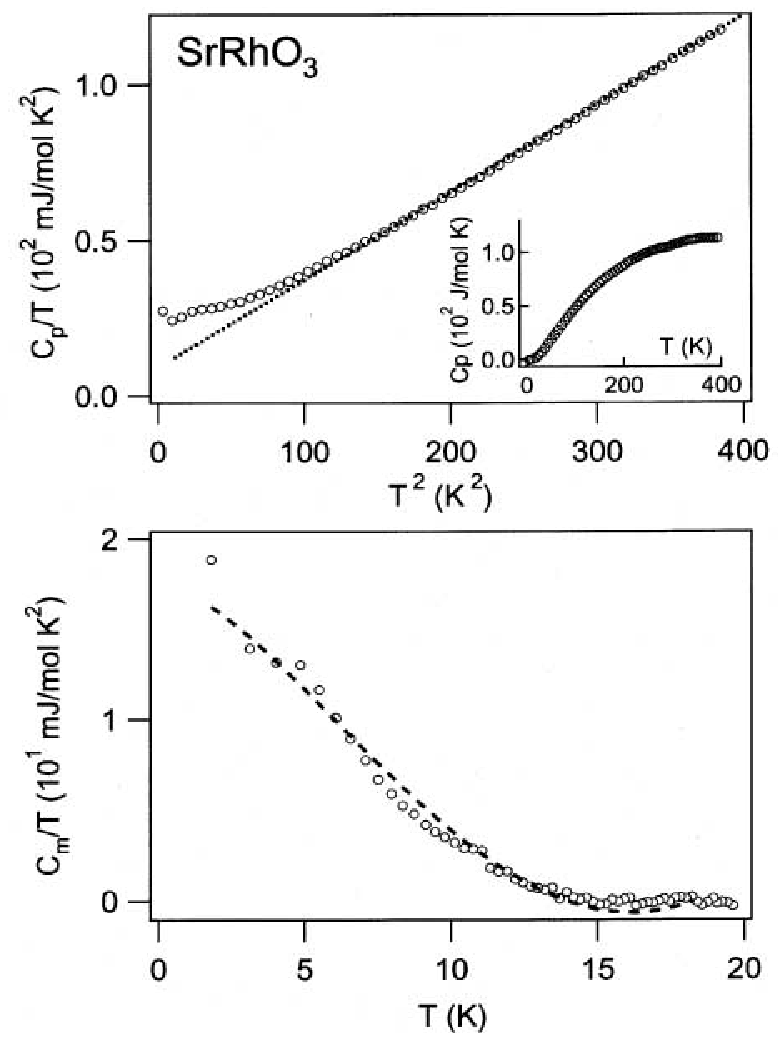}
\caption{Specific heat data for the polycrystalline SrRhO$_3$ (top) and the 
data after the orthorhombic lattice contribution and the Sommerfeld constant
are subtracted (bottom). The size of error bars are as small as the circles.
The estimated Debye temperature is 190 K and the Sommerfeld constant is 7.6 
mJ/mol K$^2$ by a least squares fitting as indicated by the dotted line. The
broken curve in the bottom panel is computed (see text).}
\label{Cp}
\end{figure*}

The specific heat data are presented in Fig.\ref{Cp}. 
A standard relaxation technique was employed in measurement. 
The temperature dependence of the specific heat ($C_{\rm p}$) of SrRhO$_3$ 
measured between 1.8 and 390 K is plotted in the inset of the top panel in 
Fig.\ref{Cp} after subtraction of a contribution from the addenda.
The difference between $C_{\rm p}$ and $C_{\rm v}$ was assumed insignificant
in the temperature range studied.
The top main panel shows a $C_{\rm p}/T$ vs $T^2$ plot of the data below 20 
K.
As expected within the Debye approximation, a linear dependence is clearly 
seen. 
The estimated Debye temperature was 190 K and the Sommerfeld constant 
was 7.6 mJ/mol K$^2$ by a least squares fitting as indicated by the
dotted line in Fig.\ref{Cp}.
Among Fermi liquid metals, a universality was found in $A/\gamma^2$ \cite 
{SSC86KK}.
A tentative calculation of $A/\gamma^2$ with the obtained parameters for 
SrRhO$_3$, $\gamma =$ 7.6 mJ/mol K$^2$ and $A =$ 0.062 $\mu\Omega$cm/K$^2$,
produced an incredible result, a value approximately two magnitudes larger than the
universal constant.
The parameter $A$ for SrRhO$_3$ perhaps involves extrinsic contributions 
somewhat as $\rho_0$ dose. 
We decided, therefore, not to make further quantitative analysis for 
$A/\gamma^2$ of SrRhO$_3$.
On the other hand, we found that the Debye temperature of SrRhO$_3$ is much 
lower than those of the ruthenium oxide perovskites \cite{PRL99KY}.
This fact would indicate the lattice of SrRhO$_3$ is much `softer' 
than that of the ruthenium oxide perovskites. 
As expected from the Debye temperature, even within the studied temperature range,
it can be clearly seen that the specific heat is approaching the roughly expected 
value $\sim$ 125 mJ/mol K [5(atoms per unit cell)$\times$3 (dimensionality per
atom) $\times k_{\rm B}$N(Boltzmann and Avogadro's constants)]. 

In the low temperature portion of the specific heat data, an extra 
contribution ($C_ {\rm m}$) appears, as $C_{\rm p}/T$ starts to part gradually
from the linear dependence on cooling.
It is presumably magnetic in origin and is found in a variety of itinerant 
magnetic materials \cite{PRL99KY,JPF83JWL}. 
The probable magnetic term was extracted by subtracting the lattice 
contribution and the Sommerfeld constant from the original data, which is shown
in the bottom panel of Fig.\ref{Cp}.
At first, the $C_{\rm m}/T$ data was quantitatively investigated with a component for spin 
fluctuations in the SCR framework for nearly ferromagnetic metals
\cite{PRL99KY,JPSJ87RK}.
The contribution of the spin fluctuations to the specific heat was approximated by  
\begin{eqnarray}
{C_{\rm m} \over T} \sim {9N_0 \over T_0}\int^{1/K_0}_0\mathrm{d}xx^2{1 \over t}
\bigg[-u-{1 \over 2}+u^2\Psi'(u) \bigg],
\label{CmT}
\end{eqnarray}
where $N_0$ is the number of magnetic atoms, $\Psi'(u)$ is the first 
derivative of the digamma function, $T_0$ and $K_0$ are the 
parameters as to spin fluctuations, $u = x(x^2+\chi(0)/\chi)/t$, $t = T/T^*$, and 
$T^* = T_0/K_0^3$ \cite{JPSJ87RK}.
The expression was then reduced to the following form in the low temperature limit:
\begin{eqnarray}
\rightarrow {3N_0 \over 4T_0}\bigg[\ln{(1+K_0^{-2})}+{2 \over 5}t^2\ln{t}
+\cdots \bigg].
\label{CmT2}
\end{eqnarray}
For fitting purposes by a least squares method, $T_0$, $K_0$, and $T^*$ 
were set as independent variable parameters in the
first two terms in Eq.\ref{CmT2}, where $t$ was replaced by $T/T^*$.
The best fitting result is shown in the bottom panel of Fig.\ref{Cp}
as a broken curve. 
Although the observed data, $C_{\rm m}/T$ vs $T$, were reproduced at a 
convincible level, all of the parameters determined here, $T_0 =$ 0.00305 K,
$K_0=$ 3.69 and $T^*=$ 26.7 K were, however, incredible \cite{JPSJ87RK}.
For example, the tentatively obtained values do not satisfy the form $T^* = 
T_0/K_0^3$ at all.
As dictated by Eq.\ref{CmT2}, there were no other combinations of the
parameters that fit the data. 
These facts, therefore, suggest that the contribution from spin fluctuations
in nearly ferromagnetic metals is either unlikely or at least insufficient to
account for the observed $C_{\rm m}$ in SrRhO$_3$. 
In 3D nearly antiferromagnetic metals, magnetic contributions to the specific heat
in the SCR framework have been studied; an enhancement of $\gamma$ is 
expected at low temperature instead of the parameters in Eq.\ref{CmT2} \cite
{PRL70TM}.
The 3D nearly antiferromagnetic picture is, therefore, unlikely to explain the
observed $C_{\rm m}$ for SrRhO$_3$.

\section{Conclusions}
The structure and electronic properties of a polycrystalline sample of 
SrRhO$_3$ obtained by high-pressure synthesis techniques was investigated.
Although the polycrystalline nature of the sample limited quantitatively detailed 
analysis of the electrical resistivity properties, the present data strongly suggests
the perovskite is in the category of a Fermi liquid.
The magnetic susceptibility of SrRhO$_3$ was found to follow a rather unusual 
temperature dependence, i.e. $1/\chi \sim T^2$. 
The tentative attempt of quantitative analysis using 3D spin fluctuation 
models resulted in inconvincible results for the magnetic susceptibility 
and the specific heat data. 
Although the major contribution to the enhancement of the paramagnetism of 
SrRhO$_3$ might result from a seizable density of state at Fermi level, as in
(Sr,Ca)RuO$_3$ \cite{PRB99JO,PRL99KY,PRB01TH}, it is not sufficient to 
explain the entire magnetic behavior of SrRhO$_3$, because it is 
temperature-independent.
There are likely additional factors which account for the temperature
dependent portion of the magnetism with the $1/\chi \sim T^2$ trend.
The character of the paramagnetism of SrRhO$_3$ seems to be intermediate between that
of enhanced Pauli- and CW-type paramagnetism.
While extensive studies were made on paramagnon contributions for the CW 
paramagnetism in the vicinity of the critical point, the intermediate
paramagnetism was essentially uninvestigated.
Whether the rhodium oxide 3D metal tends toward either an antiferro- or 
ferromagnetic instability point, the imposing appearance of the distinctive 
$T^2$ term in $1/\chi$ indicates that the magnetic excitation of 4$d$ electrons in 
SrRhO$_3$ remains highly elusive.
Further investigations into SrRhO$_3$, including theoretical consideration, would
be of significant interest.

\acknowledgments
We are grateful to Dr. D.P. Young (Louisiana State Univ.) for helpful 
discussions.
We wish to thank Drs. M. Akaishi and S. Yamaoka (AML/NIMS) for their advice 
on the high-pressure experiments.
This research was supported in part by the Multi-Core Project administrated 
by the Ministry of Education, Culture, Sports, Science and Technology of Japan.

\begin{references}
\bibitem{PT01YM}
Y. Maeno, T.M. Rice, and M. Sigrist, Phys. Today {\bf 54}, 42 (2001).
\bibitem{PRB99JO}
J. Okamoto, T. Mizokawa, A. Fujimori, I. Hase, M. Nohara, H. Takagi, Y. 
Takeda, and M. Takano, Phys. Rev. B {\bf 60}, 2281 (1999); K. Fujioka, J. Okamoto,
T. Mizokawa, A. Fujimori, I. Hase, M. Abbate, H.J. Lin, C.T. Chen, Y. Takeda, and 
M. Takano, Phys. Rev. B {\bf 56}, 6380 (1997). 
\bibitem{PRL99KY}
K. Yoshimura, T. Imai, T. Kiyama, K.R. Thurber, A.W. Hunt, and K. Kosuge, 
Phys. Rev. Lett. {\bf 83}, 4397 (1999); T. Kiyama, K. Yoshimura, K. Kosuge, 
H. Mitamura, and T. Goto, J. Phys. Soc. Jpn. {\bf 68}, 3372 (1999); T. Kiyama,
K. Yoshimura, K. Kosuge, H. Michor, and G. Hilscher, J. Phys. Soc. Jpn. {\bf
67}, 307 (1998); G. Cao, S. McCall, M. Shepard, J.E. Crow, and R.P. Guertin,
Phys. Rev. B {\bf 56}, 321 (1997).
\bibitem{PRB01TH}
T. He and R.J. Cava, Phys. Rev. B {\bf 63}, 172403 (2001); T. He, Q. Huang, 
and R.J. Cava, Phys. Rev. B {\bf 63}, 024402 (2000).
\bibitem{PRB00SII}
S.I. Ikeda, Y. Maeno, S. Nakatsuji, M. Kosaka, and Y. Uwatoko, Phys. Rev. B 
{\bf 62}, R6089 (2000).
\bibitem{binarysystem}
Rh$_2$O$_3$-SrO quasi-binary system at ambient pressure was reported 
elsewhere: R. Hory\'{n}, Z. Bukowski, M. Wo{\l}cyrz, and A.J. Zaleski, J. Alloys 
Comp. {\bf 262-263}, 267 (1997); J.R. Plaisier, A.A.C. van Vliet, and D.J.W.
Ijdo, J. Alloys Comp. {\bf 314}, 56 (2000); J.B. Claridge and H.-C. zur Loye,
Chem. Mater. {\bf 10}, 2320 (1998).
\bibitem{press}
The apparatus is a standard belt-type. Pressure and temperature were 
calibrated prior to the synthesis runs: S. Yamaoka, M. Akaishi, H. Kanda, T.
Osawa, T. Taniguchi, H. Sei, and O. Fukunaga, J. High Pressure Inst. Jpn. {\bf
30}, 249, (1992).
\bibitem{Rietveld}
F. Izumi and T. Ikeda, Mater. Sci. Forum {\bf 321-324}, 198 (2000).
\bibitem{TGA} 
A small pice ($\sim$ 15 mg) of the pellet was slowly heated up in 
mixture gas (3 \% hydrogen in argon) to 800 $^\circ$C and held until weight 
reduction was enough saturated. Calculated oxygen composition from the weight loss 
data was SrRhO$_{3.05}$, which presumed to be slightly overestimated.
\bibitem{RMP98MI}
M. Imada, A. Fujimori, and Y. Tokura, Rev. Mod. Phys. {\bf 70}, 1039 (1998).
\bibitem{JPCM96SRJ}
S.R. Julian, C. Pfleiderer, F.M. Grosche, N.D. Mathur, G.J. McMullan. A.J. 
Diver, I.R. Walker, and G.G. Lonzarich, J. Phys.:Condens. Matter {\bf 8}, 9675
(1996).
\bibitem{PRB93AJM}
A.J. Millis, Phys. Rev. B {\bf 48}, 7183 (1993). 
\bibitem{JPSJ98AI}
A. Ishigaki and T. Moriya, J. Phys. Soc. Jpn. {\bf 67}, 3924 (1998); A. 
Ishigaki and T. Moriya, J. Phys. Soc. Jpn. {\bf 65}, 3402 (1996); T. Moriya 
and T. Takimoto, J. Phys. Soc. Jpn. {\bf 64}, 960 (1995).
\bibitem{PPS61HK}
H. Kojima, R.S. Tebble, and D.E.G. Williams, Proc. Phys. Soc. {\bf A260}, 237
(1961); D.W. Budworth, F.E. Hoare, and J. Preston, Proc. Phys. Soc. {\bf A257}, 250
(1960); F.E. Hoare and J.C. Walling, Proc. Phys. Soc. {\bf B64}, 337 (1951).
\bibitem{upturn}
The major part was subtracted by employing the empirical fitting formula as 
discussed later $\chi_{\rm upturn} = [1/\chi_{\rm total}- 
(0.0026T^2+658.03)]^{-1}$.
\bibitem{preCW}
The $\chi_0$ obtained in the preliminarily fitting study was usually in 
negative $10^{-4}\sim10^{-3}$ order in emu/mol-Rh unit. A reasonable solution
for $\chi_0$ was never found.
\bibitem{Springer85TM}
T. Moriya, {\it Spin Fluctuations in Itinerant Electron Magnetism}, edited by
Manuel Cardona (Springer-Verlag, 1985); T. Moriya and K. Ueda, Adv. Phys. {\bf
49}, 555 (2000).
\bibitem{JPSJ73TM}
T. Moriya and A. Kawabata, J. Phys. Soc. Jpn. {\bf 34}, 639 (1973); T. Moriya
and A. Kawabata, J. Phys. Soc. Jpn. {\bf 35}, 669 (1973).
\bibitem{JPSJ75HH}
H. Hasegawa, J. Phys. Soc. Jpn. {\bf 38}, 107 (1975); H. Hasegawa and T. Moriya, 
J. Phys. Soc. Jpn. {\bf 36}, 1542 (1974).
\bibitem{JPSJ87RK}
R. Konno and T. Moriya, J. Phys. Soc. Jpn. {\bf 56}, 3270 (1987); K. Ueda, 
Sold State Commun. {\bf 19}, 965 (1976); K. Ueda and T. Moriya, J. Phys. Soc.
Jpn. {\bf 39}, 605 (1975).
\bibitem{RMP75KGW}
K.G. Wilson, Rev. Mod. Phys. {\bf 47}, 773 (1975). 
\bibitem{PRL68MTB}
M.T. B\'{e}al-Monod, Shang-Keng Ma, and D.R. Fredkin, Phys. Rev. Lett. {\bf 
20}, 929 (1968).
\bibitem{SSC86KK}
K. Kadowaki and S.B. Woods, Solid State Commun. {\bf 58}, 507 (1986). 
\bibitem{JPF83JWL}
J.W. Loram and Z. Chen, J. Phys. F:Met. Phys. {\bf 13}, 1519 (1983); G.R. 
Stewart, J.L. Smith, A.L. Giorgi, and Z. Fisk, Phys. Rev. B {\bf 25}, 5907 
(1982); R.J. Trainor, M.B. Brodsky, and H.V. Culbert, Phys. Rev. Lett. {\bf 
34}, 1019 (1975).
\bibitem{PRL70TM}
T. Moriya, Phys. Rev. Lett. {\bf 24}, 1433 (1970); T. Moriya, Phys. Rev. Lett. 
{\bf 25}, 197(E) (1970).

\end{references}
\end{document}